# Efficient Active Learning Strategies for Computer Experiments

Difan Song and V. Roshan Joseph[*]

H. Milton Stewart School of Industrial and Systems Engineering,
Georgia Institute of Technology, Atlanta, GA, 30332

## Abstract

Active learning in computer experiments aims at allocating resources in an intelligent manner based on the already observed data to satisfy certain objectives such as emulating or optimizing a computationally expensive function. There are two main ingredients for active learning: an initial experimental design, which helps to approximately learn the function, and a surrogate modeling technique, which provides a prediction of the output along with its uncertainty estimates. Space-filling designs are commonly used as initial design and Gaussian processes for surrogate modeling. This article aims at improving the active learning procedure by proposing a new type of initial design and a new correlation function for the Gaussian process. The ideas behind them are known in other fields such as in sensitivity analysis or in kernel theory, but they never seem to have been used for active learning in computer experiments. We show that they provide substantial improvement to the state-of-the-art methods for both emulation and optimization. We support our findings through theory and simulations, and a real experiment on the vapor-phase infiltration process.

[*]Corresponding author: roshan@gatech.edu.

# 1  Introduction

Expensive black-box functions and models appear in a wide range of applications, such as rocket engine design (Mak et al., 2018), materials informatics (Krishna et al., 2023), and instrument optimization (Knapp et al., 2023), to name a few. See the books Santner et al. (2018), Fang et al. (2005), and Gramacy (2020) for more examples. The high cost of evaluating such functions presents challenges for allocating appropriate resources.

There are usually two ways to approach the problem. The first approach is to dedicate all computational budget to a space-filling design (Joseph, 2016). Such single-batch designs aim to have points everywhere in the design region to extract information about the potentially nonlinear response surface. In contrast, the *active learning* approach uses only a proportion of the budget as the initial design. After evaluating the black-box function on this design, a surrogate model is constructed. Then, based on the predictions and the uncertainty in the surrogate model, new design points are sequentially selected using an *acquisition function*, and the model is updated. In this stage, different acquisition functions can be used for different objectives, making the procedure highly flexible. The process continues until the budget is exhausted.

Figure 1 summarizes the active learning procedure. Throughout this paper, we assume that the input variables $\mathbf{x} = (x_1, \ldots, x_d)'$ lie in the unit hypercube $[0,1]^d$. We also assume that the black-box function is evaluated without error and returns a single output $y = f(\mathbf{x})$. These input-output pairs are used for surrogate modeling. One of the most popular choices for a surrogate model is the Gaussian process (GP, Rasmussen and Williams, 2006). In a GP, given a set of $n$ inputs $\mathcal{D} = \{\mathbf{x}_i\}_{i=1}^n$, the corresponding outputs $\mathbf{y} = \{y_i\}_{i=1}^n$ have a multivariate normal distribution. By the conditional property of the normal distribution, the posterior distribution for a new input $\mathbf{x} \in \mathcal{X}$ is also normal:

$$y(\mathbf{x}) \mid \mathbf{y} \sim \mathcal{N}(\hat{y}(\mathbf{x}), s^2(\mathbf{x})),$$

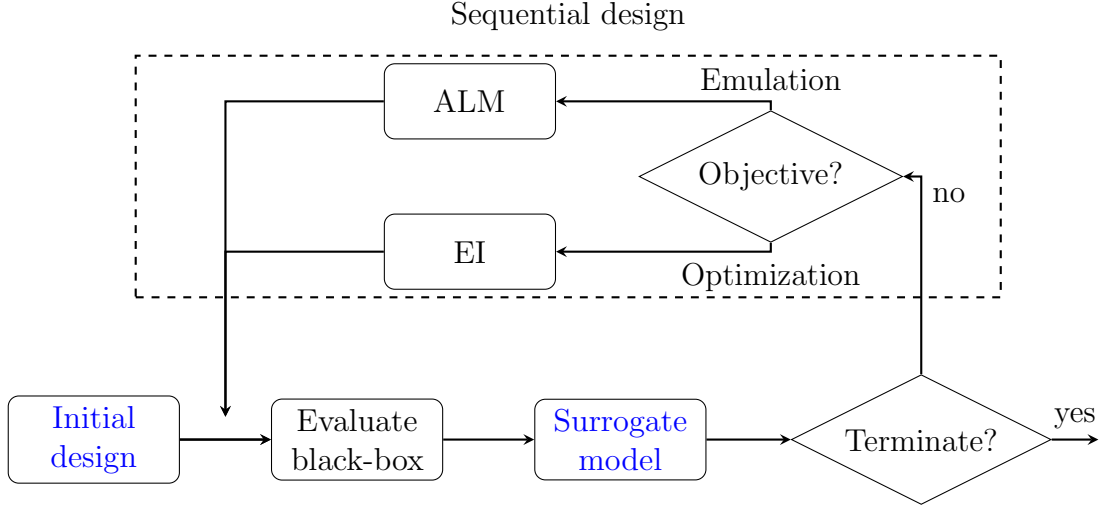

**Figure 1:** *Active learning procedure for black-box functions.*

which provides a point estimator $\hat{y}(\mathbf{x})$ and a corresponding variance $s^2(\mathbf{x})$. This property makes the GP attractive for simultaneous prediction and uncertainty quantification, enabling sequential design strategies for different purposes.

A GP is completely characterized by a mean function $\mu(\mathbf{x}) = \mathbb{E}[y(\mathbf{x})]$ and a covariance function $C(\mathbf{x}_i, \mathbf{x}_j)$. The simplest specification, known as *ordinary kriging*, assumes $\mu(\mathbf{x}) = \mu$ and $C(\mathbf{x}_i, \mathbf{x}_j) = \sigma^2 R(\mathbf{x}_i, \mathbf{x}_j)$, where $R(\cdot, \cdot)$ is a correlation kernel. By far the most widely used kernel in computer experiments is the Gaussian kernel (Santner et al., 2018):

$$R(\mathbf{x}_i, \mathbf{x}_j) = \exp\left(-\sum_{k=1}^{d} \frac{(x_{ik} - x_{jk})^2}{\theta_k^2}\right),$$

where $\theta_1, \ldots, \theta_d$ denote the length scales for the $d$ variables. Let $\mathbf{R}$ denote the correlation matrix of the current $n$ observations, and $\mathbf{r}(\mathbf{x}) = (R(\mathbf{x}, \mathbf{x}_1), \ldots, R(\mathbf{x}, \mathbf{x}_n))'$ the correlation vector. The posterior mean and variance now become:

$$\begin{aligned}
\hat{y}(\mathbf{x}) &= \mu + \mathbf{r}(\mathbf{x})'\mathbf{R}^{-1}(\mathbf{y} - \mu\mathbf{1}), \\
s^2(\mathbf{x}) &= \sigma^2 \left\{1 - \mathbf{r}(\mathbf{x})'\mathbf{R}^{-1}\mathbf{r}(\mathbf{x})\right\},
\end{aligned} \quad (1)$$

where $\mu$, $\sigma^2$, and $\theta_1, \ldots, \theta_d$ are estimated using empirical Bayes methods (Currin et al., 1991).

When sequentially learning about a black-box function, there are often two distinct objectives: emulation and optimization. Emulation refers to building a surrogate model to approximate the function with sufficient accuracy (Sacks et al., 1989), which usually involves the entire input space. On the other hand, optimization aims to find the specific set of inputs that minimizes (or maximizes) the output. For emulation, a simple but effective acquisition is called Active Learning MacKay (ALM, MacKay, 1992; Gramacy, 2020). The ALM criterion selects the point with the largest posterior variance, which is a well-known criterion used in optimal experimental designs (Fedorov, 1972). The criterion suggests that we should devote our resources to where we are most uncertain, which is very intuitive. In a GP, the points with the largest variance are those that are the "furthest" from the nearest design point, so this criterion ensures that we are thoroughly exploring the design space. On the other hand, during optimization, the acquisition should reflect the exploration-exploitation trade-off. When using a GP surrogate, the process is also known as Bayesian optimization (BO). There are many such acquisition functions (Frazier, 2018; Garnett, 2023), and the most popular choices include the expected improvement (Jones et al., 1998), upper confidence bounds (Wang et al., 2016), and knowledge gradient (Frazier et al., 2008). All of these choices strike a balance between the predicted value and the associated uncertainty.

Despite the wide application and success of active learning with GP surrogates, the method has limitations when applied to black-box functions with many inputs. As a motivating example, consider the Levy function evaluated on $[-10, 10]^d$, which contains many local optima and is commonly used for testing optimization algorithms:

$$f(\mathbf{x}) = \sin^2(\pi w_1) + \sum_{k=1}^{d-1}(w_k - 1)^2 \left[1 + 10\sin^2(\pi w_k + 1)\right] + (w_d - 1)^2 \left[1 + \sin^2(2\pi w_d)\right],$$

$$\text{where } w_k = 1 + \frac{x_k - 1}{4}, \text{ for all } k = 1, \ldots, d.$$

The first issue concerns the initial design stage. As previously mentioned, space-fillingness has been the main focus for such single-batch designs. Common choices that "cover" the experimental region well include the maximin and the minimax design (Johnson et al., 1990). To increase the one-dimensional projections, Morris and Mitchell (1995) proposed to use a maximin Latin hypercube design (MmLHD) that maximizes the minimum distance among the design points. Joseph et al. (2015b) further proposed the MaxPro design that ensures good projections on all subspaces of the inputs. MmLHD and MaxPro designs remain the most popular choices. From the perspective of better estimating the length scale parameters of a surrogate model, Zhang et al. (2021) proposed to consider the distribution of the pairwise distances of the design points. To this end, they proposed beta-distributed designs integrated with space-filling designs (betadist) to improve model fitting.

However, we point out that, while the aforementioned designs are good choices for estimating the response surface, none of them put an emphasis on identifying important variables. In many black-box functions that simulate a real process, there are many input variables, but only a few are influential on the output. Successfully identifying the influential inputs during the initial design stage allows us to allocate the budget more efficiently during the later stages.

Consider an example of a 10-dimensional function where the first six dimensions follow the Levy function, and the last four dimensions are inert. We use betadist or MaxPro designs of size $10d = 100$ as initial designs, and the length scales estimated on the initial designs are plotted. The design size follows the rule-of-thumb suggested by Loeppky et al. (2009). However, as shown in Figure 2, for both designs, the estimated length scales do not show clear evidence of which dimensions are important. Therefore, we cannot expect the active learning procedure to devote more resources to the important variables.

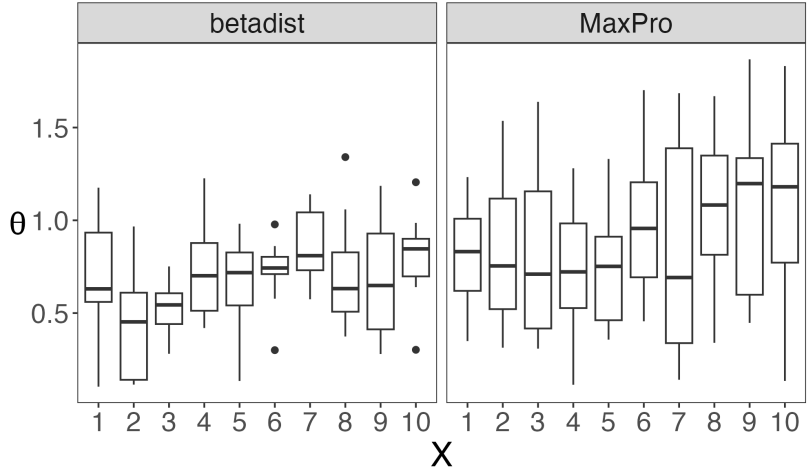

**Figure 2:** *Boxplots of the estimated GP length scale parameters of a 10-dimensional Levy function. The first six dimensions are active, and the last four dimensions are inert. For both the betadist and the MaxPro designs, the design size is $10d = 100$ points. The estimation is repeated for ten random seeds.*

The second issue concerns low-dimensional projections in the sequential design stage. For both emulation and optimization, we would like to explore the design space with good projections. When the domain is $\mathcal{X} = [0, 1]^d$, the design should ideally fill the interval $[0, 1]$ for each dimension. While we can easily achieve this using MmLHD or MaxPro in the initial design, it is not at all easy in the sequential exploration stage (Xu et al., 2015). Therefore, we face the risk of missing low-dimensional signals in particular regions of the design space during the sequential stage.

Figure 3 visualizes the points selected with ALM during the emulation of the Levy function, where the black dots are the sequential design. For $d = 2$, the selected points are reasonable. However, for $d = 6$, we observe a tendency to overpopulate the boundary. Consequently, the low-dimensional projections become very poor, and much of the information in the interior will be missing.

Existing approaches address the boundary issue by either modifying the acquisition function or incorporating prior information that discourages boundary solutions. For emulation, Gramacy (2020) suggests using the active learning Cohn (ALC) criterion, which focuses on the *integrated* variance over the entire design space rather than the variance of a single point. For optimization, Pourmohamad and Lee (2022) uses a barrier function approach, which is a natural strategy in constrained optimization. Their resulting acquisition, named "one over sigma squared" (OOSS), favors interior points when there is limited information. Siivola et al. (2018) instead proposes augmenting virtual derivative sign observations on the boundary to drive points away.

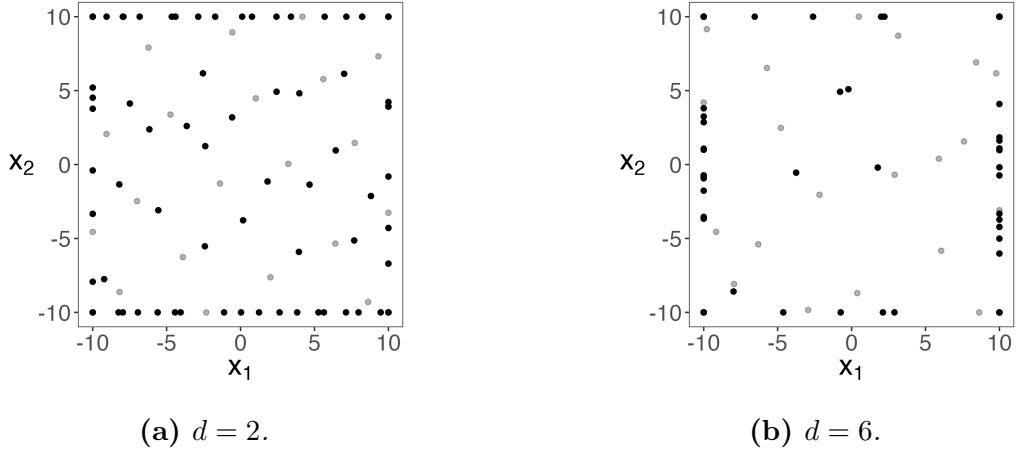

(a) $d = 2$.

(b) $d = 6$.

**Figure 3:** *Emulation of the Levy function with the Gaussian kernel. The gray dots show the initial design (20 points) and the black dots show the sequentially selected points (80 points). For $d = 6$, the projection onto the first two dimensions is shown.*

In this work, we propose to tackle the two aforementioned issues from design and modeling perspectives. Our procedure uses a specialized screening design named *maximum one-factor-at-a-time* (MOFAT) design (Xiao et al., 2023) and a specialized correlation kernel named the *multiplicative inverse multiquadric* (MIM) kernel. Specifically, we will show that the MOFAT design efficiently identifies important input dimensions with a small design size in the initial stage, while the MIM kernel automatically incorporates the MaxPro criterion (Joseph et al., 2015b) in the sequential stage. The components involved in the procedure are present in the literature, but their roles in active learning have been underexplored.

The remaining parts of the paper are outlined as follows. Section 2 introduces the procedure, including details on the MOFAT design and the MIM kernel. Theoretical results are also included in this section. Section 3 compares simulation performance on a suite of test functions for both emulation and optimization. Section 4 applies the procedure to a model calibration application. We conclude with some remarks in Section 5.

## 2 Methodology

In this section, we propose two improvements to the active learning strategy in computer experiments.

### 2.1 Screening designs

Instead of space-filling and distance-based designs, we propose using screening designs in the initial design stage of active learning. Although popular in physical experiments, to the best of our knowledge, such designs have not been used for active learning purposes in computer experiments.

Identifying a relatively small number of important variables from a large number of variables has always been an important topic in the design of experiments (DOE) literature. By the effect sparsity principle (Wu and Hamada, 2021), focusing on the few factors vital to the outcome is desirable. In the case of physical experiments, famous examples include the Plackett-Burman designs (Plackett and Burman, 1946) and the supersaturated design (Lin, 1993) that attempt to identify the significant main effects of two-level factors using as few runs as possible. For three-level factors, the definitive screening designs (Jones and Nachtsheim, 2011) allow the estimation of quadratic effects in addition to linear effects.

However, for computer experiments or black-box functions with continuous inputs, the nonlinear effects and complex interactions are often of interest. Therefore, we need specialized screening designs for computer experiments for which the Morris screening design (Morris, 1991) is a popular choice. The main idea behind Morris screening design is to use a collection of one-factor-at-a-time (OFAT) designs that are randomly placed in the experimental region. Each OFAT design consists of $d+1$ runs (because of $d$ changes from a base run) and there are $l$ such OFAT designs. This leads to a total of $l(d+1)$ runs for the entire design. Each OFAT design gives an estimate of the local sensitivity of the variables. By aggregating the local sensitivities obtained from the $l$ randomly distributed OFAT designs, we can get a global sensitivity index of each variable, which can be used to decide whether the variable is active or not.

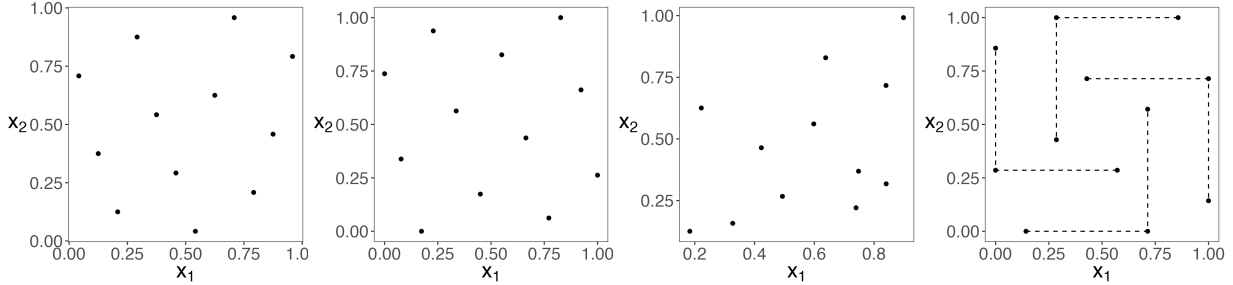

| (a) Maximin LHD. | (b) MaxPro Design. | (c) Betadist design. | (d) MOFAT Design. |

**Figure 4:** *Visualization of several initial designs for $d = 2$.*

Xiao et al. (2023) recently made a connection between Morris screening designs and the pick-and-freeze Monte Carlo-based designs for estimating the total Sobol' indices (Sobol', 1993). They found that Sobol's Monte Carlo designs are also OFAT designs but with random perturbations. By exploiting this connection, Xiao et al. (2023) optimized the OFAT structure of Sobol's Monte Carlo designs to improve the efficiency of the screening. They named their designs as Maximum one-factor-at-a-time (MOFAT) designs.

Figure 4 presents several 12-point initial designs in two dimensions. The MmLHD, MaxPro, and betadist designs have good one-dimensional projection properties in the sense that none of the design points have duplicate values for a single variable. However, this characteristic also complicates identifying the causal effects from the input variables to the output. In contrast, the OFAT structure in MOFAT makes it straightforward to attribute variation in the outcome to each variable, although we pay the price of slightly worse projections. Furthermore, the plot shows $l = 4$, which means four base points are varied in $d = 2$ dimensions to get a design size of $l(d+1)$. For larger $d$, the choice of $l = 4$ offers a great reduction in the design size compared with the $10d$ rule. Therefore, we can allocate more resources to the sequential design stage and exploit the knowledge in the initial design. We will show through numerical examples that a screening design of this size can work better than a large space-filling design.

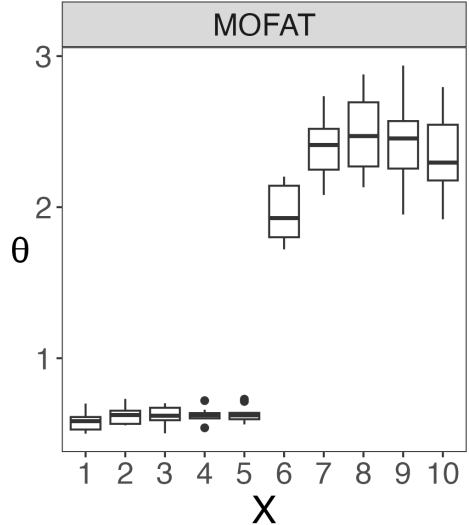

**Figure 5:** *Boxplot of the estimated GP length scale parameters of a 10-dimensional Levy function when a MOFAT design is used. The first six dimensions are active, and the last four dimensions are inert. The design size is $l(d+1) = 44$ points. Estimation is repeated for ten random seeds.*

Figure 5 revisits the 10-dimensional Levy function example, and again the first six dimensions are active. We fit a GP with the Gaussian kernel based on a 44-point ($l = 4$) MOFAT design. The length scales now show clear evidence that the first five dimensions are the most influential to the output (recall that the smaller the length scale, the faster the correlation decays). The sixth dimension, while active, is not as important. Therefore, by using a screening design as the initial design, we can identify impactful input variables with much less budget, which can potentially lead to improvements in emulation and optimization. We will investigate this more using simulations in Section 3.

## 2.2 The MIM kernel

The second component of our procedure is a new correlation function for the Gaussian process:

$$R(\mathbf{x}_i, \mathbf{x}_j) = \prod_{k=1}^{d} \left(1 + \frac{(x_{ik} - x_{jk})^2}{\theta_k^2}\right)^{-\alpha}, \alpha > 0. \tag{2}$$

This kernel is the product of one-dimensional inverse multiquadric (IM) kernels (Hardy and Göpfert, 1975); hence we refer to it as the multiplicative inverse multiquadric (MIM) kernel. Note that the commonly used IM kernel in higher dimensions has the form (Fasshauer, 2007, p.41):

$$R(\mathbf{x}_i, \mathbf{x}_j) = \left(1 + \frac{\|\mathbf{x}_i - \mathbf{x}_j\|_2^2}{\theta^2}\right)^{-\alpha}, \quad \alpha > 0 \tag{3}$$

and its anisotropic version is given by $R(\mathbf{x}_i, \mathbf{x}_j) = \left\{1 + \sum_{k=1}^{d}(x_{ik} - x_{jk})^2/\theta_k^2\right\}^{-\alpha}$, which is quite different from (2). The IM kernel is readily available in many software packages: `scikit-learn` (Pedregosa et al., 2011), `pykernels` (Czarnecki and Janocha, 2016), and `gpytorch` (Gardner et al., 2018) in Python, `KSPM` (Schramm et al., 2020) in R, to name a few. On the other hand, the MIM kernel has not received much attention in the literature with the exception of Mak and Joseph (2017). They referred to this kernel as the sparsity-inducing (SpIn) kernel and used it for high-dimensional data reduction purposes. Here, we highlight some features of this kernel that are relevant to surrogate modeling.

First, the MIM kernel is better at detecting low-dimensional structures. Through the product form, the kernel provides a low-dimensional similarity measure for high-dimensional points. Consequently, the kernel is more efficient in detecting the main effects of the variables, which are probably the most significant effects in complex models by the effect hierarchy principle (Wu and Hamada, 2021). We expect the MIM to perform better than the Gaussian kernel for functions with many input variables and low-dimensional structures.

Second, the MIM kernel emphasizes the differences in *individual coordinates*. Specifically, when $\theta_k$ is close to zero, only the points with $x_{ik} \approx x_{jk}$ will have a high correlation. Compared to the IM kernel 3, the MIM kernel avoids being a function of the Euclidean distance when $d \geq 2$. In Section 1, we discussed the limitation associated with kernels that depend on the Euclidean distance. Here, we make this intuition more precise by analyzing the variance of the ordinary kriging predictor.

Our first result is with respect to the IM kernel (3). In the spirit of Johnson et al. (1990), we investigate the case where $\theta \to 0$, i.e., near-independence of the Gaussian process. The proof of the theorems in this section can be found in Appendix A.

**Theorem 1.** *Consider fitting a Gaussian process (ordinary kriging) with the IM kernel* (3) *and the design* $\mathcal{D} = \{\mathbf{x}_1, \ldots, \mathbf{x}_n\}$. *Then, for all* $\mathbf{x} \notin \mathcal{D}$:

$$\lim_{\theta \to 0} \frac{\sigma^2 - s^2(\mathbf{x})}{\theta^{4\alpha} \sigma^2} = \sum_{i=1}^{n} \frac{1}{\|\mathbf{x} - \mathbf{x}_i\|^{4\alpha}}.$$

Thus, for small $\theta$, maximizing $s^2(\mathbf{x})$ is equivalent to minimizing $\sum_{i=1}^{n} 1/\|\mathbf{x} - \mathbf{x}_i\|^{4\alpha}$. Thus, the ALM becomes

$$\mathbf{x}_{n+1} = \operatorname*{argmin}_{\mathbf{x}} \sum_{i=1}^{n} \frac{1}{\|\mathbf{x} - \mathbf{x}_i\|^{4\alpha}},$$

which can be viewed as the sequential version of a maximin-type design (Joseph et al., 2015a). Thus, Theorem 1 shows a direct connection between the design acquired by ALM and the maximin design. If Latin hypercube restrictions are not imposed, the sequential design will have unfavorable one-dimensional projections. The same issue exists for the Gaussian kernel and the Matérn kernel, although some approximations are needed to arrive at the maximin criterion (Johnson et al., 1990). This analysis confirms the limitation of sequential designs with commonly used kernels.

We now turn to the MIM kernel and apply the same analysis for $\theta_k \to 0$, for all $k = 1, \ldots, d$.

**Theorem 2.** *Consider fitting a Gaussian process (ordinary kriging) with the MIM kernel (2) and the design $\mathcal{D} = \{\mathbf{x}_1, \ldots, \mathbf{x}_n\}$. Then, for all $\mathbf{x} \notin \mathcal{D}$:*

$$\lim_{\{\theta_k\}_{k=1}^d \to 0} \frac{\sigma^2 - s^2(\mathbf{x})}{\sigma^2} \prod_{k=1}^d \frac{1}{\theta_k^{4\alpha}} = \sum_{i=1}^n \frac{1}{\prod_{k=1}^d |x_l - x_{ik}|^{4\alpha}}.$$

Theorem 2 shows that in low-correlation regimes, ALM leads to the sequential MaxPro design (Joseph, 2016):

$$\mathbf{x}_{n+1} = \underset{\mathbf{x}}{\operatorname{argmin}} \sum_{i=1}^n \frac{1}{\prod_{k=1}^d |x_k - x_{ik}|^{4\alpha}}.$$

By inspection, for any new point with a coordinate the same as an existing point, the criterion will go to infinity. Thus, good projection properties are automatically enforced in sequential design. This avoids the need to use Latin hypercube constraints, which are difficult to incorporate into one-point-at-a-time sequential designs (Xu et al., 2015; Ba et al., 2018).

We can also see how the automatic relevance determination (ARD) mechanism works in the non-limiting case. The denominators become $\theta_k^2 + (x_k - x_{ik})^2$, indicating that the difference $x_k - x_{ik}$ is less impactful on the criterion when $\theta_k$ becomes large. The parameters $\theta_1, \ldots, \theta_d$ identify the variables for which the output is "wiggly". Although the length scale parameters are not appropriate sensitivity measures (Paananen et al., 2019), they are the correct parameters to consider for sequential designs. For small $\theta_k$, the design will put effort into the projection of the corresponding $x_k$; conversely, a large $\theta_k$ does not necessarily mean the variable is insignificant for output prediction, but the smoothness indicates that a design towards the boundary will be more informative.

A comparison between Theorems 1 and 2 suggests that the multiplicative form emphasizes individual coordinates, leading to better projections. It is natural to also consider whether multiplicative forms work for other kernels. Clearly, a multiplicative form makes no difference for the Gaussian kernel. As for the Matérn kernel, we consider a special case, the Matérn 1/2 kernel (also known as the exponential kernel): $R(x_i, x_j) = \exp(-|x_i - x_j|/\theta)$. Instead of the Euclidean distance, the multiplicative form in multiple dimensions is now a function of $\sum_{k=1}^{d} |x_{ik} - x_{jk}|$, which is the Manhattan distance between $\mathbf{x}_i$ and $\mathbf{x}_j$, also suffering from the boundary issue. For other choices, such as Matérn 3/2 and Matérn 5/2, we can also observe a similar dependence on the Manhattan distance. Therefore, the connection to the MaxPro design is a unique property of the MIM kernel, and we can rely on this kernel to systematically explore the design space.

We now return to the emulation of the Levy function and plot the selected points using the MIM kernel in Figure 6. Compared with Figure 3, which uses the Gaussian kernel, the projections of the design in the active dimensions in the six-dimensional case have improved greatly. On the other hand, for the inert dimensions, the points are concentrated on the boundary. In this case, the effective sample size increases for active inputs, potentially leading to a better understanding of the effects.

## 3 Numerical experiments

In this section, we evaluate the performance of our proposed procedure (MOFAT design + MIM kernel) against existing settings (MaxPro design + Gaussian kernel) for various

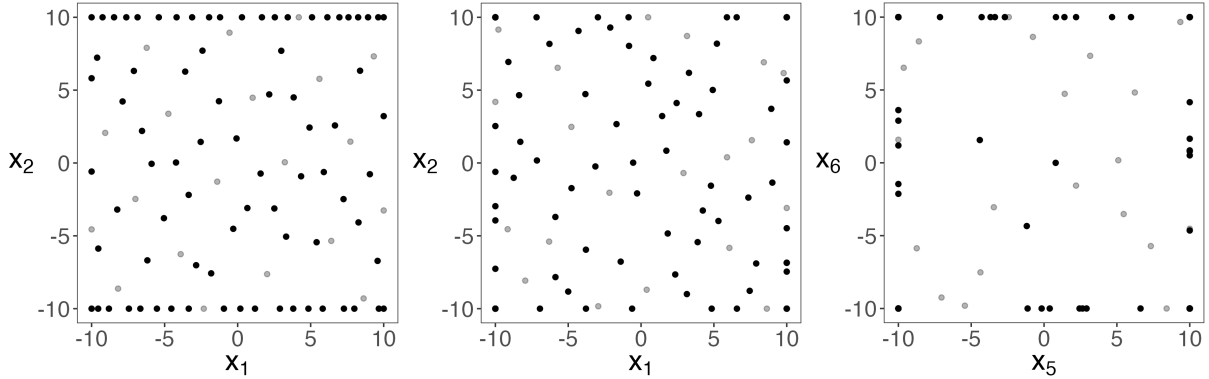

**(a)** $d = 2$.  **(b)** $d = 6$, *active dimensions.*  **(c)** $d = 6$, *inert dimensions.*

**Figure 6:** *Emulation of the Levy function with the MIM kernel. The gray dots show the initial design (20 points) and the black dots show the sequentially selected points (80 points). For $d = 6$, the first four dimensions are active, and panel (b) shows the projection onto the first two dimensions; the last two dimensions are inert, and panel (c) shows the projection onto these two dimensions.*

functions. The MOFAT and the MaxPro designs are generated using the R packages `MOFAT` (Xiao and Joseph, 2022) and `MaxPro` (Ba and Joseph, 2018). In all simulations, the parameters $\mu, \sigma^2, \alpha$ and $\theta_1, \ldots, \theta_d$ are estimated by maximum likelihood estimation. The length scale parameters $\theta_1, \ldots, \theta_d$ have an initial value of 0.1. The parameter $\alpha$ in the MIM kernel has an initial value of 1 and is bounded in the interval $[0.01, 3]$. All kernels have a nugget term $\eta = 10^{-6}$ to avoid numerical issues. The test functions are selected from Surjanovic and Bingham (2015). Details of all the functions are presented in the supplementary material.

## 3.1 Emulation

For emulation, we consider two sets of functions. The first two functions (the Dette & Pepelyshev function and the Friedman function) are numerical examples; the other four (the OTL circuit function, the piston simulation function, the robot arm function, and the wing weight function) are developed from physical models. The original input dimensions of the functions range from 5 to 10. For each function, we augment $10 - d$ inert variables so that the total dimension is 10.

Following the procedure in 1 for each function, we use either the MaxPro design as the initial design and fit GPs with the Gaussian kernel, or the MOFAT design as the initial design and fit GPs with the MIM kernel. The size of the initial designs is $4(10 + 1) = 44$, and we use the ALM criterion in both models to acquire a sequential design until $n = 100$.

We evaluate the out-of-sample mean squared error (MSE) on a 1000-point Sobol' sequence for each test function and show the results in Figure 7. We also generate a 100-point MaxPro design, build surrogates using both the Gaussian and the MIM kernel, and evaluate the MSE on the same Sobol' sequence. The errors provide a baseline for how different kernels perform with a batch design and are plotted as horizontal dashed lines in Figure 7. Our simulations show that the two kernels have comparable performance on all the problems.

For active learning, we first observe that the Gaussian kernel fails to reach the same level of performance as the batch design. As discussed in previous sections, too much exploration of the boundary is the main reason. On the other hand, MOFAT + MIM always achieves improved performance compared with the batch design. Consequently, the procedure outperforms Gauss + MaxPro in all test functions (except the piston simulation function, where all procedures are very accurate). This shows that, by identifying the important input dimensions and being projection-aware, our procedure explores the design space more efficiently in the sequential stage.

## 3.2 Optimization

For the numerical experiments of BO, we consider three functions with many local optima: the Ackley function, the Levy function, and the Rastrigin function. For each function, we consider two settings. In the first setting, we have six-dimensional inputs that are all active. In the second setting, we add four inert variables to make the optimization

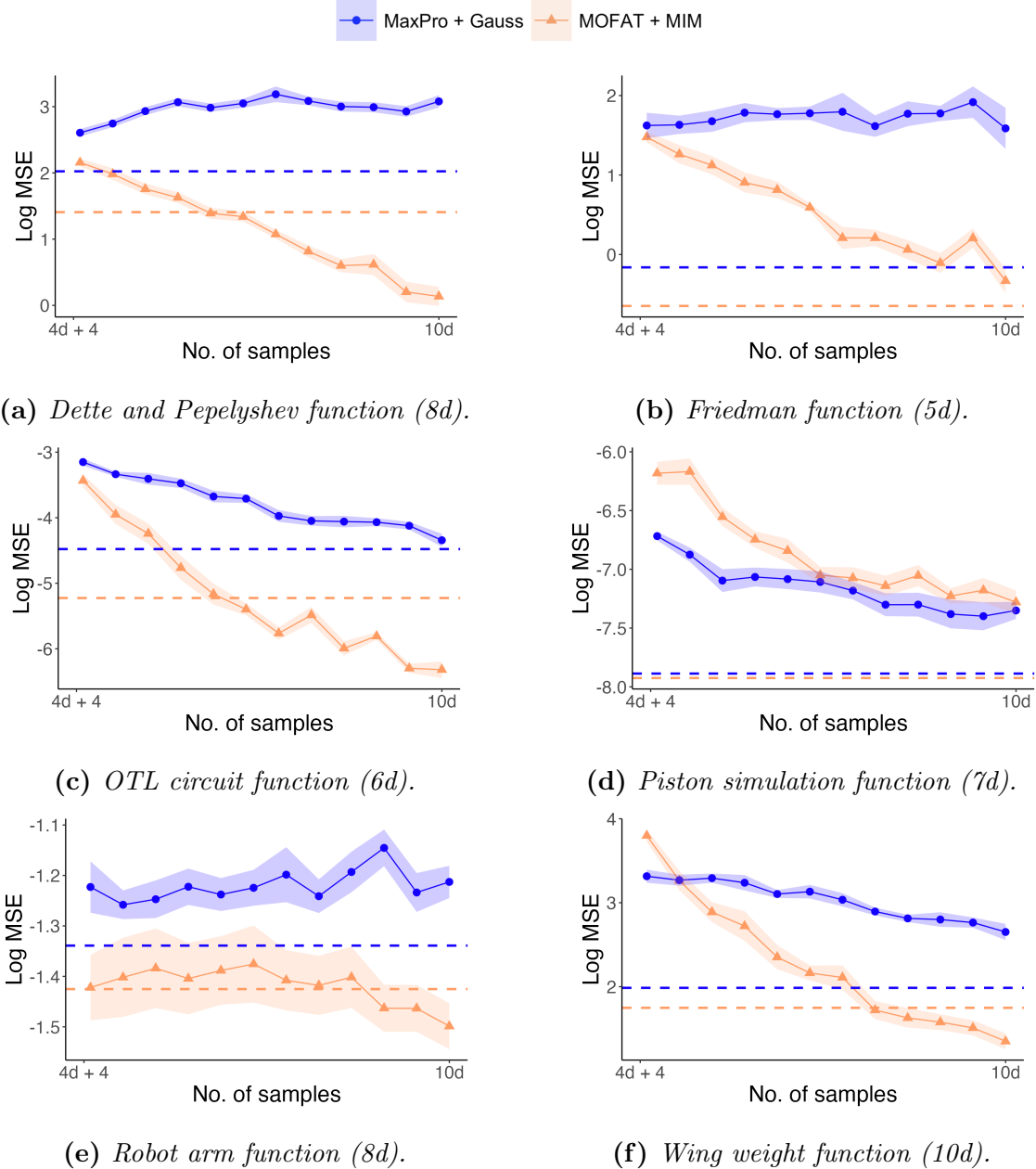

**(a)** *Dette and Pepelyshev function (8d).*

**(b)** *Friedman function (5d).*

**(c)** *OTL circuit function (6d).*

**(d)** *Piston simulation function (7d).*

**(e)** *Robot arm function (8d).*

**(f)** *Wing weight function (10d).*

**Figure 7:** *Out-of-sample mean squared errors (MSE) with fully sequentially selected samples. For all functions except the wing weight function, the input is augmented to $d = 10$ dimensions to include inert variables. In all plots, the solid lines represent the out-of-sample errors, and the horizontal dashed lines represent the errors achieved by a 100-point MaxPro design. All results are averaged over 10 runs.*

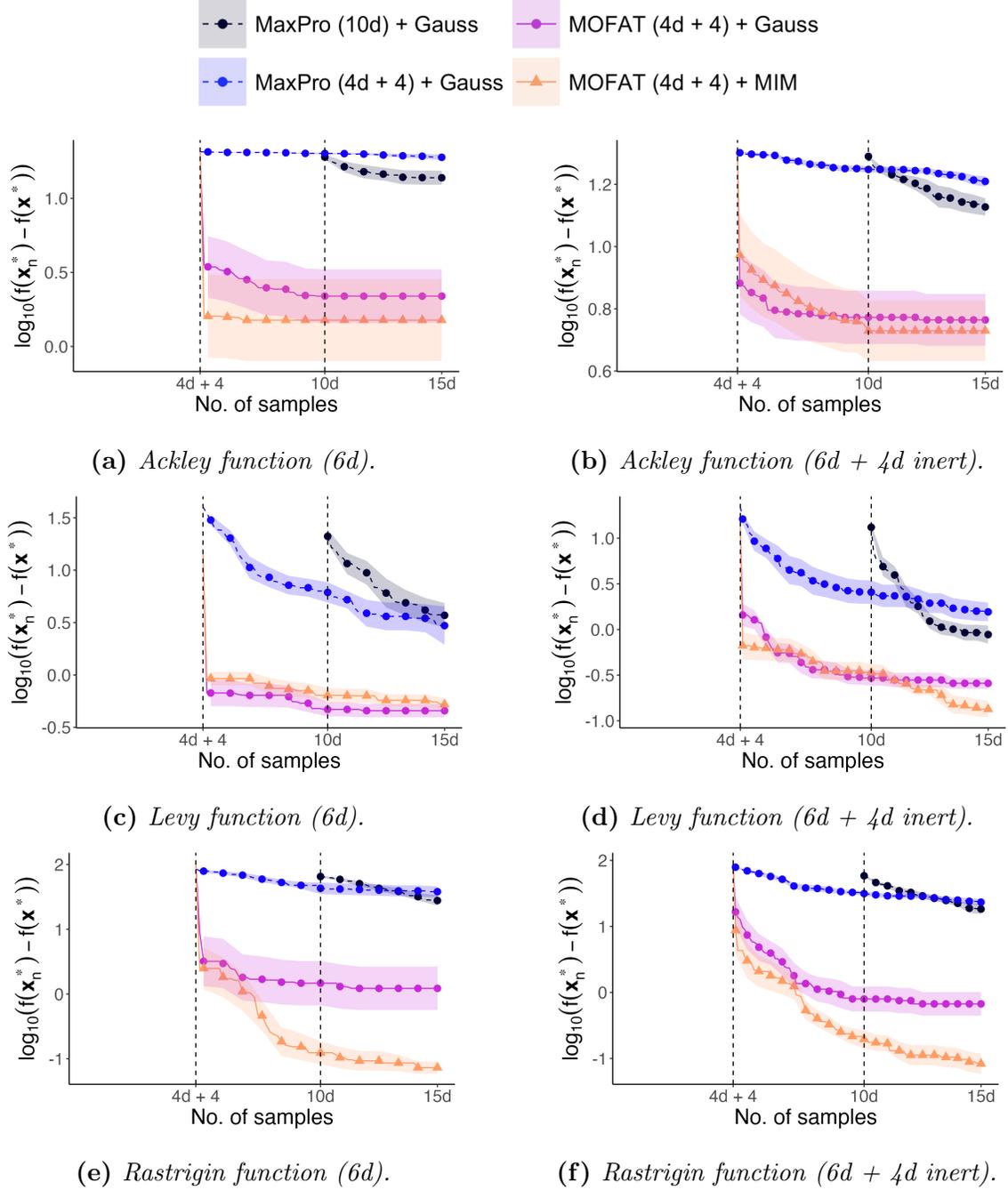

**Figure 8:** *Optimality gaps for several functions. For all examples, the size of the initial design is either $n_{init} = 4(d+1)$ or $n_{init} = 10d$, as shown in parentheses of the legends. All results are averaged over 10 runs.*

problem more difficult. For all settings, we apply our procedure with a MOFAT design with $4(d+1)$ as the initial design, and use the MIM kernel in the GP surrogate. We also use MaxPro designs of sizes $4(d+1)$ and $10d$, and GP surrogates with the Gaussian kernel for comparison.

We use the EI criterion (Jones et al., 1998) as the acquisition function:

$$\text{EI}(\mathbf{x}) = \mathbb{E}[\max\{y_n^* - \hat{y}(\mathbf{x}), 0\}],$$

where we assume the problem is minimization, and $y_n^* = \min\{y_i\}_{i=1}^n$ denotes the best value so far. Under the normal posterior, the EI criterion has a closed-form expression

$$\text{EI}(\mathbf{x}) = \{y_n^* - \hat{y}(\mathbf{x})\}\Phi(u(\mathbf{x})) + s(\mathbf{x})\phi(u(\mathbf{x})), \text{where } u(\mathbf{x}) = \frac{y_n^* - \hat{y}(\mathbf{x})}{s(\mathbf{x})},$$

where $\Phi(\cdot)$ and $\phi(\cdot)$ denote the distribution and density functions of the standard normal, respectively. Maximization of the EI criterion would lead to selecting points that either have a small predicted value (exploitation) or large uncertainty (exploration). We continue the process until depleting a budget of $n = 15d$.

In Figure 8, we present the optimality gaps. First, we notice that for the MaxPro design, reducing the design size from $10d$ to $4d + 4$ will lead to a deterioration of the optimization performance. Indicating that a smaller sample size is insufficient for a space-filling design to explore the design space. In contrast, the MOFAT design with size $4d + 4$ offers dramatic improvements over MaxPro designs. In all examples, active learning can exploit the information from the MOFAT design and lead to immediate improvements in the first few sequential design points. Finally, by using the MIM kernel, we can exploit the information from the screening design even better, especially when there are low-dimensional structures in the black-box function. Therefore, we can conclude that both in emulation and optimization, our procedure is more successful at handling a large number of inputs relative to the budget, leading to potential savings in computational resources and improved efficiency.

# 4 Application: parameter identification in the VPI process

In this section, we present an application where we use BO to calibrate the unknown parameters in a system of partial differential equations (PDEs). Vapor phase infiltration (VPI) is an emerging process for producing organic-inorganic hybrid materials with potential applications in a number of commercial industries (Leng and Losego, 2017). To understand the fundamental thermodynamic and kinetic properties associated with the process, Ren et al. (2021) developed a reaction-diffusion model governed by the following PDEs:

$$\begin{cases} \frac{\partial C_{\text{free}}}{\partial t} = D \frac{\partial^2 C_{\text{free}}}{\partial s^2} - k C_{\text{free}} C_{\text{polymer}} \\ \frac{\partial C_{\text{product}}}{\partial t} = k C_{\text{free}} C_{\text{polymer}} \\ D = D_0 \exp\left(-K' C_{\text{product}}\right) \\ \frac{\partial C_{\text{polymer}}}{\partial t} = -k C_{\text{free}} C_{\text{polymer}} \end{cases}$$

with the boundary and initial conditions:

$$\begin{cases} C_{\text{free}} = 0, & 0 < s < L, t = 0 \\ C_{\text{product}} = 0, & 0 < s < L, t = 0 \\ C_{\text{polymer}} = C_{\text{polymer}}^0, & 0 < s < L, t = 0 \\ \frac{\partial C_{\text{free}}}{\partial s} = 0, & s = 0, t > 0 \\ C_{\text{free}} = C_s, & x = L, t > 0. \end{cases}$$

**Table 1:** *Basic information of the calibration parameters in the PDEs. The fourth column provides the estimated total Sobol' indices from the initial MOFAT design. The fifth column is the optimal values found by our procedure with the MOFAT design and the MIM kernel.*

| Notation | Description | Range | $\hat{t}$ | Final value |
|---|---|---|---|---|
| $D_0$ | Diffusivity | $[10^{-12}, 10^{-9}]$ | 0.3663 | $2.640 \times 10^{-10}$ |
| $C_s$ | Surface concentration | $[0.004, 0.005]$ | 0.0285 | $4.206 \times 10^{-3}$ |
| $C_0$ | Accessible polymer concentration | $[0.005, 0.006]$ | 0.0129 | $5.808 \times 10^{-3}$ |
| $K'$ | Hindering | $[500, 2500]$ | 0.4182 | 1214 |
| $k$ | Reaction rate | $[0.001, 10]$ | 0.4178 | 1.390 |

In these equations, $C_{\text{free}}$ (mol/cm$^3$) is the concentration of the free diffusing vapor-phase precursor, $C_{\text{polymer}}$ (mol/cm$^3$) is the concentration of the accessible reactive polymeric functional groups, $C_{\text{product}}$ (mol/cm$^3$) is the concentration of immobilized product from the reaction between the free-diffusing vapor-phase precursor and the polymeric functional groups. There are five unknown parameters $\mathbf{x} = \{D_0, C_s, C_0, K', k\}$ that directly impact the process, where $D_0$ (cm$^2$/s) is the initial diffusivity of the free diffusing vapor-phase precursor, $C_s$ (mol/cm$^3$) is the surface concentration of the free diffusing vapor-phase precursor, $C_0$ (mol/cm$^3$) is the initial concentration of accessible reactive polymeric functional groups, $K'$ (cm$^3$/mol) is the hindering factor describing how immobilized product $C_{\text{product}}$ slows down the diffusivity of free diffusing vapor, and $k$ (cm$^3$/mol·s) is the associated reaction rate. In Table 1, we present the input domains for these parameters. See Ren et al. (2021) for more details on the PDEs and the VPI process.

Under experimental conditions, the total mass uptake of the free diffusing vapor-phase precursor and the immobilized product (i.e., $C_{\text{free}} + C_{\text{product}}$) over period $[0, 120000]$ (seconds) is collected. Our objective is to calibrate the parameters $\mathbf{x} = \{D_0, C_s, C_0, K', k\}$ so that the PDE solutions match the real data as close as possible in terms of the MSE:

$$f(\mathbf{x}) = \int_0^T \{H(t; \mathbf{x}) - m(t)\}^2 dt,$$

where $H(t; \mathbf{x})$ denotes the PDE outputs of $C_{\text{free}} + C_{\text{product}}$ that depends on the parameters $\mathbf{x}$, $m(t)$ denotes the experimental data, and $T = 120,000$ seconds. The experimental data and PDE implementation are available at `https://github.com/BillHuang01/BOFO`, and we focus on the 8.7 Torr experimental data in this work.

For each set of $\mathbf{x}$, solving the PDE to get $H(t; \mathbf{x})$ takes about one minute on a 2.3 GHz laptop. Hence, BO is suitable for the task of identifying the best parameters within a limited number of runs. Huang et al. (2021) applied a functional principal component analysis (fPCA) approach to take advantage of the functional output nature of $H(t; \mathbf{x})$ and $m(t)$. In their results, matching the first three component scores from fPCA led to a better set of parameters than directly applying BO on the MSE or the logarithm of the MSE. In this work, we still apply BO on the logarithm of the MSE but use the proposed MOFAT design and the MIM kernel to see how much of an improvement the procedure can offer.

For the five-dimensional optimization problem, Huang et al. (2021) used a MaxPro design with $10d = 50$ points as the initial design, and used active learning to select 25 more points. With MOFAT and $l = 4$, our initial design contains 24 points. With this design, we can apply the method of Oakley and O'Hagan (2004), and estimate the total Sobol' indices from the GP surrogate using the MIM kernel. The fourth column of Table 1 presents the estimated values, and we can see that $D_0, K', k$ are more influential.

The smallest root mean squared errors (RMSE) during optimization is shown in panel (a) of Figure 9. Guided by the information provided by MOFAT, we can find a parameter setting with a small RMSE ($\approx 700$) with either the Gaussian or the MIM kernel. Notice that this value is below the horizontal line, which is the RMSE of the setting found in Huang et al. (2021) ($\approx 800$). We also observe that, by using the MIM kernel, we are able to arrive at a good setting with less than 40 evaluations of the PDE system, while the Gaussian kernel requires more budget to arrive at the optimum. The final parameter settings found by the MOFAT design and the MIM kernel are presented in the last column of Table 1,

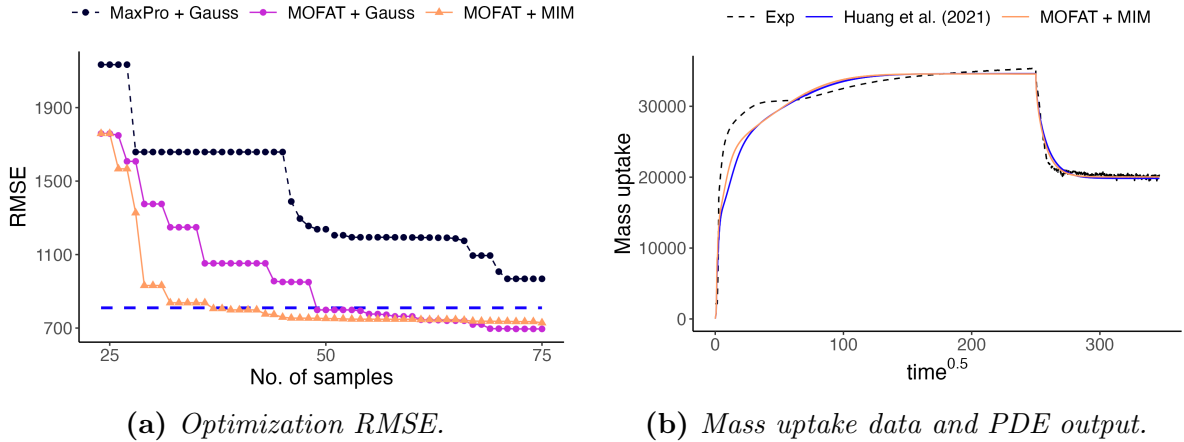

**(a)** *Optimization RMSE.*  **(b)** *Mass uptake data and PDE output.*

**Figure 9:** *(a) Minimum RMSE achieved during the calibration of the VPI process model. The horizontal dashed line indicates the minimum RMSE found by Huang et al. (2021). (b) Visualization of the mass uptake data reported in Ren et al. (2021). The other two lines plot the best fit from Huang et al. (2021) and our procedure with the MOFAT design and the MIM kernel, respectively.*

which has significant differences in the three influential parameters $D_0, K', k$ compared with Huang et al. (2021). Finally, we plot the experimental data and the outputs from Huang et al. (2021) and our procedure in panel (b) of Figure 9. The PDE outputs are very similar, and the improvement in RMSE mainly comes from larger uptake values in the early process.

Overall, we can conclude that for this five-dimensional optimization problem, our procedure with the MOFAT design and the MIM kernel is able to efficiently identify the best parameters with a small computational budget. We believe the procedure can be useful for active learning in a broad range of science and engineering applications.

# 5 Conclusions

This article proposes two improvements to active learning: MOFAT design as an initial design and MIM kernel as the correlation function in Gaussian processes. The idea of using a screening design is well known in the field of physical experiments, but popular screening designs such as Plackett-Burman designs and Definitive Screening Designs are not deemed useful in computer experiments because of their poor projections. Morris screening designs are well-known in the field of sensitivity analysis, but they are not used in computer experiments because of their poor space-filling property. Similarly, the idea of an MIM kernel is known in the theory of kernels, but no one has ever used it in practice. In this article, we bring together these ideas for active learning in computer experiments. Surprisingly, they brought substantial improvements to existing methods. We provided a theoretical justification for the use of MIM kernels as it leads to MaxPro-type designs instead of maximin-type designs. In high-dimensional problems where sparsity is present, this provides a great improvement in the way the points are spread out. Moreover, although MOFAT does not look space-filling in the full-dimensional space, it can quickly cut off the space of inactive variables, which seems to be the right strategy for starting the active learning procedure.

There exist many directions for future research. For example, the current procedure works best for problems with a moderate number of input dimensions. It will be interesting to investigate how screening designs and the MIM kernel can be used for variable selection in high-dimensional problems. Another direction is to combine the approach with nonstationary models such as deep GP (Sauer et al., 2023) to obtain non-uniform designs, fully exploiting the benefits of active learning. Finally, although we demonstrated the approach with the ALM and EI acquisitions, it can be combined with other acquisitions such as HEI (Chen et al., 2023). Investigating the theoretical properties and empirical performances of such a comprehensive procedure can provide important insights for active learning.

# Acknowledgments

This research is supported by U.S. National Science Foundation grant DMS-2310637.

# A   Proofs of Theorems 1 and 2

*Proof.* From (1), we obtain $\{\sigma^2 - s^2(\mathbf{x})\}/\sigma^2 = \mathbf{r}(\mathbf{x})'\mathbf{R}^{-1}\mathbf{r}(\mathbf{x})$. For this quadratic form, we have the relationship

$$\frac{1}{\lambda_n}\mathbf{r}(\mathbf{x})'\mathbf{r}(\mathbf{x}) \leq \mathbf{r}(\mathbf{x})'\mathbf{R}^{-1}\mathbf{r}(\mathbf{x}) \leq \frac{1}{\lambda_1}\mathbf{r}(\mathbf{x})'\mathbf{r}(\mathbf{x}), \tag{4}$$

where $\lambda_1, \lambda_n$ denote the smallest and largest eigenvalues of $\mathbf{R}$, respectively. When $\mathbf{x} \notin \mathcal{D}$, for the IM kernel, we have the limit

$$\lim_{\theta \to 0} \frac{1}{\theta^{4\alpha}} \mathbf{r}(\mathbf{x})'\mathbf{r}(\mathbf{x}) = \lim_{\theta \to 0} \sum_{i=1}^{n} \frac{1}{(\theta^2 + \|\mathbf{x} - \mathbf{x}_i\|^2)^{2\alpha}} = \sum_{i=1}^{n} \frac{1}{\|\mathbf{x} - \mathbf{x}_i\|^{4\alpha}}.$$

Furthermore, $\mathbf{R} \to \mathbf{I}$ indicates $\lambda_1, \lambda_n \to 1$. Now the result in Theorem 1 follows from (4) and the squeeze theorem.

For the MIM kernel, (4) still holds. Similarly, consider the limit

$$\lim_{\{\theta_l\}_{l=1}^d \to 0} \mathbf{r}(\mathbf{x})'\mathbf{r}(\mathbf{x}) \prod_{l=1}^{d} \frac{1}{\theta_l^{4\alpha}} = \lim_{\{\theta_l\}_{l=1}^d \to 0} \sum_{i=1}^{n} \frac{1}{\prod_{l=1}^{d}\{\theta_l^2 + (x_l - x_{il})^2\}^{2\alpha}}$$

$$= \sum_{i=1}^{n} \frac{1}{\prod_{l=1}^{d} |x_l - x_{il}|^{4\alpha}},$$

With $\lambda_1, \lambda_n \to 1$, the result in Theorem 2 follows from the squeeze theorem. $\square$